\documentclass[12pt, preprint]{aastex}

\usepackage{graphicx}
\usepackage{amssymb}
\usepackage{layout}
\usepackage{url}
\usepackage{subfigure}
\usepackage{color}
\usepackage{enumerate}
\slugcomment{Submitted to {\it The Astrophysical Journal}}

\begin{document}


\title{Radiation and Polarization Signatures of 3D Multi-zone Time-dependent Hadronic Blazar Model}


\author{Haocheng Zhang\altaffilmark{1,2}, Chris Diltz\altaffilmark{3}, and Markus B\"ottcher\altaffilmark{4}}

\altaffiltext{1}{Department of Physics and Astronomy, University of New Mexico, Albuquerque, NM 87131, USA}

\altaffiltext{2}{Theoretical Division, Los Alamos National Laboratory, Los Alamos, NM 87545, USA}

\altaffiltext{3}{Astrophysical Institute, Department of Physics and Astronomy, Ohio University, Athens, OH 45701, USA}

\altaffiltext{4}{Centre for Space Research, North-West University, Potchefstroom,
2520, South Africa}

\begin{abstract}
We present a newly developed time-dependent three-dimensional multi-zone hadronic blazar emission model. 
By coupling a Fokker-Planck based lepto-hadronic particle evolution code 3DHad with a polarization-dependent radiation 
transfer code, 3DPol, we are able to study the time-dependent radiation and polarization signatures of a hadronic 
blazar model for the first time. Our current code is limited to parameter regimes in which the hadronic 
$\gamma$-ray output is dominated by proton synchrotron emission, neglecting pion production. Our results
demonstrate that the time-dependent flux and polarization signatures are generally dominated by the relation 
between the synchrotron cooling and the light crossing time scale, which is largely independent of the exact 
model parameters. We find that unlike the low-energy polarization signatures, which can vary rapidly in time, 
the high-energy polarization signatures appear stable. As a result, future high-energy polarimeters may be 
able to distinguish such signatures from the lower and more rapidly variable polarization signatures expected
in leptonic models.
\end{abstract}

\section{Introduction}

Blazars are the most violent class of active galactic nuclei. Their emission is known to be nonthermal-dominated, 
covering the entire electromagnetic spectrum from radio up to TeV $\gamma$-rays, with strong variability on all 
time scales \citep[e.g.,][]{Aharonian07}. Blazar spectral energy distributions (SEDs) are characterized by two
broad, non-thermal components. The low-energy component, from radio to optical-UV, is generally agreed to be 
synchrotron radiation of ultrarelativistic electrons. The origin of the high-energy component, from X-rays to 
$\gamma$-rays, is still under debate. The leptonic model argues that the high-energy component is due to the 
inverse Compton scattering of either the low-energy synchrotron emission \citep[SSC, e.g.][]{Marscher85,Maraschi92} 
or external photon fields \citep[EC, e.g.,][]{Dermer92,Sikora94}, while the hadronic model suggests that the 
high-energy emission is dominated by synchrotron emission of ultrarelativistic protons and the cascading secondary 
particles resulting from photo-pion and photo-pair production processes \citep[e.g.,][]{Mannheim92,Mucke01}. 
It is of high importance to many aspects of high energy astrophysics to distinguish these two models, because 
it will put strong constraints on the blazar jet power, the physics of the central black hole, the origin of 
ultra-high-energy (UHE) cosmic rays and very-high-energy (VHE, i.e., TeV -- PeV) neutrinos. However, both models 
are generally able to produce reasonable fits to snap-shot SEDs of blazars \citep[e.g.,][]{Boettcher13}. Thus, 
additional diagnostics are necessary.

An obvious choice would be through the identification of blazars as the sources of VHE neutrinos, which are the 
``smoking gun'' of hadronic interactions \citep[e.g.,][]{Halzen97,Kistler14,Diltz15,Petropoulou15}. IceCube has 
reported detection of astrophysical VHE neutrinos, and there are hints that the origin of these neutrinos could
be spatially connected to blazars \citep[e.g.,][]{Aartsen13,Kadler16}. However, in view of the low angular 
resolution of IceCube, so far the sources of these neutrinos are still unknown. 

An alternative is the study of light curves. The development of time-dependent leptonic models has been quite 
fruitful \citep[e.g.,][]{Joshi11,Diltz14,Weidinger15,Asano15}. Although one-zone leptonic models sometimes 
have difficulty in explaining the frequently seen symmetric light curves, some multi-zone leptonic models 
that explicitly include the light travel time effects (LTTEs) have successfully resolved that issue 
\citep[e.g.,][]{Chen14}. On the other hand, due to the more complicated cascading processes, hadronic 
models are generally stationary and/or single-zone \citep[e.g.,][]{Mastichiadis95,Cerruti15,Yan15}. 

Another possible discriminant is that leptonic and hadronic models require very distinct magnetic field 
conditions. Radio to optical polarization measurements have been a standard probe of the jet magnetic field. 
In particular, recent observations of $\gamma$-ray flares with optical polarization angle (PA) swings and 
substantial polarization degree (PD) variations indicate the active role of the magnetic field during 
flares \citep[e.g.,][]{Marscher08,Abdo10,Blinov15}. Several models have been put forward to explain 
these phenomena \citep[e.g.,][]{Larionov13,Marscher14,ZHC15}, and a first-principle magnetohydrodynamics 
(MHD) based model is also under development \citep{ZHC16}. For the high energy emission, \cite{ZHC13} have 
shown that by combining the infrared/optical and the X-ray/$\gamma$-ray polarization signatures, it would be
possible to distinguish the two models. Several X-ray and $\gamma$-ray polarimeters are currently proposed
and/or under development \citep[e.g.,][]{Hunter14}. However, despite remarkable progress that has been made 
to improve these high-energy polarimeters, they commonly suffer from limited sensitivity. If the high-energy 
polarization signatures vary as rapidly as the low-energy (optical) polarization, it will be difficult for 
these polarimeters to measure, as they will integrate over episodes of vastly different PAs. This prompted
us to investigate the time-dependent high-energy polarization signatures of lepto-hadronic blazar models in
more detail.

In this paper, we present a newly developed 3D multi-zone time-dependent hadronic model code, 3DHad. This new code is based on the one-zone time-dependent Fokker-Planck (FP) based lepto-hadronic code of \cite{Diltz15}, but generalized to 3D multi-zone. By coupling with the 3D 
polarization-dependent ray-tracing routines of the 3DPol code developed by \cite{ZHC14}, we will derive 
the time-dependent radiation and polarization signatures across the whole blazar SED, including all LTTEs. 
Hence, we can study the general phenomenology of the light curves and time-dependent polarization signatures. 
Hadronic models generally require very high jet powers and magnetic fields. Therefore, we will put physical 
constraints on the allowed parameter space by estimating the available jet power and magnetic field in the
case of a Blandford-Znajek \citep{BZ77} powered jet. With the above consideration, 
we will predict detailed time-dependent 
polarization signatures from proton-synchrotron dominated hadronic models based on various jet conditions and 
flaring mechanisms. These results can be compared with multiwavelength light curves and future high-energy 
polarization measurements, putting stringent constraints on the blazar jet conditions in a hadronic model. 
We will describe our code setup and physical considerations in Section \ref{code}, sketch our model setup 
in Section \ref{model}, present case  studies in Sections \ref{result1} and \ref{result2}, and discuss the 
results in Section \ref{discussion}.

\section{3DHad and Physical Considerations \label{code}}

In this section, we will first introduce 3DHad and its main features, and how it is coupled with 3DPol. 
Then we will justify the physical considerations for the hadronic model and put constraints on the parameter 
space. Finally, for code verification purposes, we will compare 3DHad with the one-zone hadronic code developed 
by \cite{Diltz15}, and illustrate the similarities and differences.

\subsection{Code Features}

3DHad is a time-dependent multi-zone nonthermal electron and proton evolution code based on FP equations. The code is written in a module-oriented style in FORTRAN 95, fully parallelized by MPI. This code can directly take inputs of each zone, including geometry, magnetic field information, particle evolution, etc., either from inhomogeneous blazar model parameters, or from first principle simulations such as MHD and particle-in-cell (PIC) simulations. Based on the inputs, each zone will solve FP equations for the evolution of electron and proton energy distributions. In this first application that we present here, we will not consider any particle transfer between the zones, hence the FP equations in each zone are independent.

We apply an implicit Euler method to solve the FP equations numerically. The solutions of FP equations in each zone are based on the work by \cite{Diltz15}. The general form of the 
FP equation is
\begin{equation}
\frac{\partial n(\gamma,t)}{\partial t} = \frac{\partial}{\partial \gamma} (K\gamma^2 
\frac{\partial n(\gamma,t)}{\partial \gamma}) - \frac{\partial}{\partial \gamma}((\dot{\gamma}+
2K\gamma)n(\gamma,t))-\frac{n(\gamma,t)}{t_{esc}}+n_{inj}(\gamma,t)
\label{fpeq}
\end{equation}
where $K=1/(2t_{acc})$. The underlying assumption in Eq. \ref{fpeq} is that the particle evolution is governed by four processes, a fast first-order Fermi acceleration process characterized by $n_{inj}(\gamma,t)$, a second-order Fermi acceleration process characterized by a mass-independent acceleration time scale $t_{acc}$, the synchrotron cooling on $\dot{\gamma}$, and particle escape parameterized by an energy-independent escape time scale $t_{esc}$. In a steady, quiescent state, nonthermal particles are continuously injected into each zone with a power-law distribution in energy,
\begin{equation}
n_{inj}(\gamma,t)=n_0\times\gamma^{-p},~\gamma_{min}<\gamma<\gamma_{max} 
\end{equation}
which represents a rapid particle acceleration mechanism, such as diffusive shock acceleration and magnetic reconnection \citep{Guo16} on time scales much shorter than the time resolution of our simulation. In addition to this process, the emission region may contain microscopic turbulence, which will mediate stochastic second-order Fermi acceleration. We take $t_{acc}=1/\alpha$ as the stochastic acceleration time scale, where $\alpha=\frac{d\gamma}{dt}/\gamma$ is the stochastic acceleration rate. The exact form of $\alpha$ depends on the turbulent acceleration model and the turbulence parameters. A detailed treatment of these aspects is beyond the scope of this paper. Here we simply take $t_{acc}$ to be independent of particle energy. In general, the Compton cooling rates
in hadronic models are negligible compared to synchrotron losses due to the large magnetic fields \citep[e.g.,][]{Boettcher13}, especially 
in the parameter space that we will employ here. Finally, since we do not consider particle transport between zones in this first application, we simply use an energy-independent escape time scale $t_{esc}$ to mimic the process that particles leave a particular zone and no longer contribute to the emission there.

While the original one-zone hadronic code in \cite{Diltz15} 
is very comprehensive, including all details of
pion production, $\gamma$-$\gamma$ interactions, explicit muon and pion evolution, etc., in this paper, we
will restrict the parameters to a regime in which the proton energy losses and radiative outputs are strongly
dominated by proton synchrotron emission, thus neglecting photo-pion and photo-pair production processes, 
and following only the electron and proton evolution. In this way, the photon transfer between each zone will not affect the particle evolution. The parameter restrictions inherent in this assumption will be detailed in the next section. 

3DHad solves the FP equations for the particle distributions in each zone at each time step; in order to calculate the resulting emission, the derived time-dependent particle distributions will be fed into 3DPol \citep{ZHC14}, which has been upgraded to include 
synchrotron emission (and their polarization signatures) from heavier particles, such as protons. 3DPol can calculate the time-dependent radiation and polarization signatures based on the particle and magnetic field inputs. Since in the parameter regime adopted here, Compton scattering is negligible, radiation transfer between zones will not affect the particle evolution. Hence, the radiation transfer problem is reduced to a ray-tracing method. The gyroradius of a proton is given by
\begin{equation}
r_g=\frac{\gamma_p m_pc^2}{eB} \sim 3\times 10^6 \frac{\gamma_p}{B (G)}~cm
\end{equation}
For a magnetic field of $B \sim 10$~G, and the most energetic protons around $\gamma\sim10^9$, this yields a 
gyroradius of the order of $\sim 10^{14}$~cm, which is smaller than our spatial resolution in the emission 
region. Therefore, we can assume that all particles will radiate in their corresponding zones, and calculate the 
individual Stokes parameters in each zone. By adding up the ones that arrive at the observer at the same 
time, we naturally include all LTTEs.

The execution of the combined 3DHad and 3DPol is efficient. For the runs that we will show in this paper, 
they take about 20 minutes on 500 CPUs on LANL clusters. Therefore, the code has the potential to do larger 
runs for more detailed physical modeling, e.g., with physical conditions derived from MHD simulations.

\subsection{Physical Constraints and Assumptions \label{physics}}

Hadronic blazar models usually require high magnetic fields and nonthermal particle energies close to the 
upper limits that blazars can plausibly provide based on our current understanding of accretion and jet 
formation processes \citep[e.g.,][]{Boettcher13,Cerruti15,Zdziarski15}. Here we estimate the resulting 
limits and put constraints on the hadronic model parameter space. We employ the conservation of magnetic 
flux in the jet to estimate the available magnetic field in the emission region, and the Eddington luminosity 
to constrain the total particle energy. The magnetic flux from the Blandford-Znajek mechanism is given by
\begin{equation}
\Phi_h \sim 1.4\times 10^{33}\frac{1}{f_{\Omega}(a)}L_{46}M_9 ~ G\, cm^2
\end{equation}
where $f_{\Omega}(a)=a/(1+\sqrt{1-a^2})$ and $a$ is the dimensionless spin parameter of the black hole, 
$L_{46}$ is the magnetic jet luminosity in units of $10^{46}~erg\,s^{-1}$, and $M_9$ is the black hole 
mass in units of $10^9$ solar masses. Assuming conservation of the poloidal magnetic flux along the jet, 
and that the poloidal component is comparable to the toroidal component, the magnetic flux in the emission 
region is approximated by $B\,\pi R^2$, where $R = 10^{16} \, R_{16}$~cm is the radius of the emission 
region. Given a bright blazar ($L_{46} \sim 100$) and a large central black hole mass ($M_9 \sim 1$), 
we find the first constraint,
\begin{equation}
B\times R^2 \lesssim 10^{33}~G\,cm^2
\label{constraint1}
\end{equation}
or $B \lesssim 10 \, R_{16}^{-2}$~G. Assuming bulk motion of the blazar emission region with a Lorentz 
factor $\Gamma \gg 1$, the kinetic luminosity in protons is evaluated by
\begin{equation}
	L_p \sim \pi R^2 \Gamma^2 c\, u_p
\end{equation}
where $u_p=m_p c^2 \int_1^{\infty} d\gamma n_p (\gamma) \gamma$ is the proton energy density in the rest
frame of the emission region, and $n_p(\gamma)$ is the proton spectral number density in that frame. We 
expect that the total proton kinetic luminosity should not exceed the Eddington luminosity, which is given by
\begin{equation}
L_{\rm Edd} = 4\pi GMm_pc/\sigma_T \sim 1.2\times 10^{47} M_9~erg\,s^{-1}
\end{equation}
With a bulk Lorentz factor $\Gamma$ of a few tens, and using the same black hole mass as in Eq. \ref{constraint1}, 
we obtain the second constraint,
\begin{equation}
u_p \times R^2 \lesssim 10^{33}~erg \, cm^{-1}
\label{constraint2}
\end{equation}
In spite of significant uncertainties in constraints on physical conditions in blazars, Eqs. \ref{constraint1} 
and \ref{constraint2} allow us to put some stringent constraints on parameters to be used for our models. For 
instance, hadronic models typically require magnetic fields exceeding $10~G$ \citep[e.g.,][]{Boettcher13,Cerruti15}. 
Hence by Eq. \ref{constraint1}, the size of the emission region in the comoving frame should generally be 
smaller than $10^{16}~cm$, which can be translated to a flare duration of $\sim 10~h$ in the observer's 
frame, assuming a typical Lorentz factor $\Gamma \sim 20$. Therefore, flares that last several days, in 
particular in the low-energy bands such as optical, where the LTTEs generally dominate, are unlikely to 
be of hadronic origin, unless some general physical conditions are varying on longer time scales. We will 
demonstrate this point in Sections \ref{result1} and \ref{result2}.

\begin{figure}[ht]
\centering
\includegraphics[width=\linewidth]{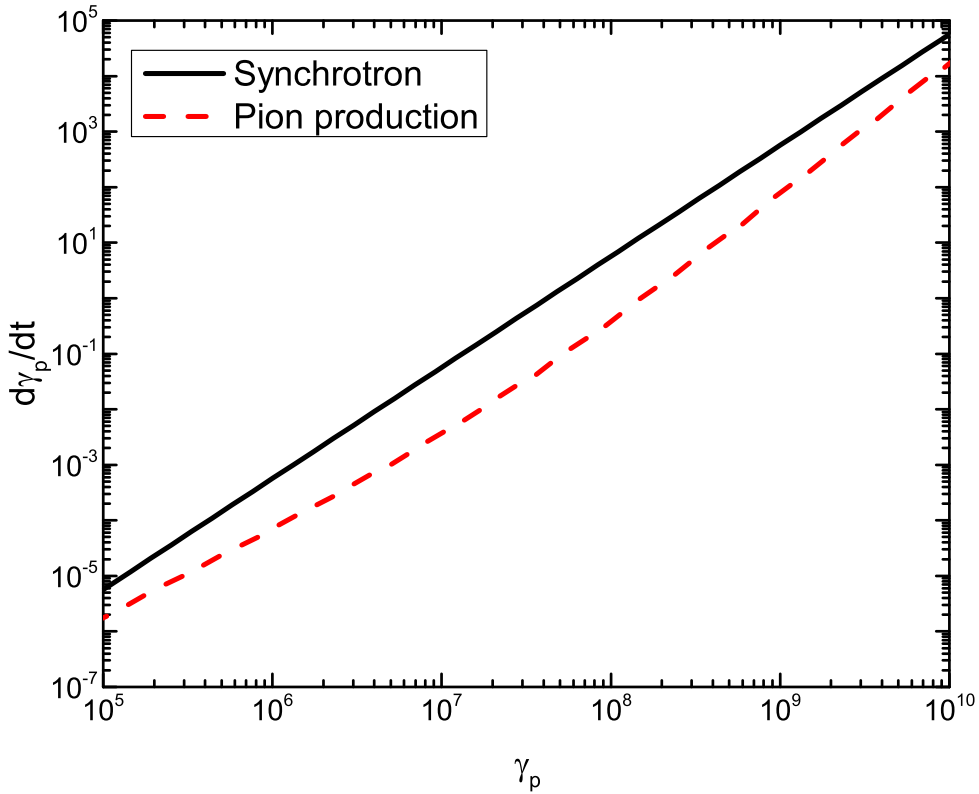}
\caption{Synchrotron and pion-production loss rates of a power law proton distribution with cut-offs 
$\gamma_{min} = 1.0$, $\gamma_{max} = 10^{8}$, spectral index $p_p = 2.2$ in an emission region of 
size $R = 10^{16}~cm$ and magnetic field of $B = 50~G$. Synchrotron losses generally dominate over 
pion-production losses for ultrarelativistic protons.
\label{assumption2}}
\end{figure}

For this preliminary study, we will make some additional assumptions in order to avoid more complicated 
radiation-feedback calculations, so that only the electron and proton evolution are important. The assumptions 
are:
\begin{enumerate}
\item Proton and electron energy losses are dominated by synchrotron cooling;
\item $\gamma\gamma$ opacity and pair-production are negligible;
\end{enumerate}
This restricts the parameter space in which our model is applicable, as detailed in the following.

For assumption 1, we need to make sure that the synchrotron cooling for protons should be faster than the 
pion-production cooling rate. The synchrotron loss rate for protons is given by
\begin{equation}
    \dot \gamma_{p,syn} = -\frac{c \sigma_{T} B^{2}}{6 \pi m_{e} c^{2}} \ (\frac{m_{e}}{m_{p}})^{3} \gamma_{p}^{2}
\end{equation}
and the pion production loss rate is given by \citep{Aharonian00}
\begin{equation}
    \dot \gamma_{p,p\gamma} = -c \langle \sigma_{p\gamma} f \rangle n_{ph} (\epsilon^{*}) \epsilon^{*} \gamma_{p}
\end{equation}
where $\langle \sigma_{p\gamma} f \rangle \sim 10^{-28}~cm^{2}$ represents the elasticity-weighted $p\gamma$ 
interaction cross section, $\epsilon^{*} = 5.9 \times 10^{-8}~E_{19}^{-1}$ represents the energy of 
target photons interacting with protons of energy $E = 10^{19} E_{19}$~eV at the $\Delta$ resonance, and 
$n_{ph} (\epsilon)$ represents the target photon field for photo-pion production in units of photon energy 
normalized with respect to the rest mass of the electron, $\epsilon = h\nu/m_{e}c^{2}$. For a typical set of
parameters of a hadronic blazar model, the two energy loss rates are plotted in Fig. \ref{assumption2}. 
By comparing the two rates, we find that synchrotron losses dominate for protons with Lorentz factors
\begin{equation}
    \gamma_{p} > \frac{6 \pi m_{e} c^{2} \langle \sigma_{p\gamma} f \rangle n_{ph} (\epsilon^{*}) 
    \epsilon^{*}}{\sigma_{T} B^{2}} \ (\frac{m_{p}}{m_{e}})^{3}
\end{equation}
Assuming that the relevant section of the target synchrotron photon spectrum in the comoving frame is in the 
form of a power-law,  $n_{ph} (\epsilon) = n_{ph}^{0} \epsilon^{-\alpha}$, the above constraint can be written 
as
\begin{equation}
    \gamma_{p} > \frac{6 \pi m_{e} c^{2} \langle \sigma_{p\gamma} f \rangle n_{ph}^{0} 
    (\epsilon^{*})^{1 - \alpha}}{\sigma_{T} B^{2}} \ (\frac{m_{p}}{m_{e}})^{3}
\end{equation}
For the highest energy protons typically used in the lepto-hadronic modeling of FSRQs, 
$\gamma_{p, max} \sim 10^{8}$, the most efficient target photons for pion production have
energies of $\epsilon^{*} = 590/\gamma_{p} \sim 6.0 \times 10^{-6}$.  This is generally in the
optical and UV bands, which is dominated by electron synchrotron emission. Using the delta 
approximation for the synchrotron power of electrons \citep{Boettcher12} and assuming a steady 
state electron distribution in the form of a power law, the constraint can then be rewritten in 
terms of the model parameters,
\begin{equation}
(10^{-17}B(G)\gamma_p)^{\frac{p_e-3}{2}}R(cm)n_{0,e}(cm^{-3})<10^{22}
\end{equation}
where $n_{0,e}$ is the normalization factor of the electron distribution, $p_e$ is the electron 
power-law index, $R$ is the radius of the emission region, and $\gamma_p$ is the proton Lorentz factor. 
As we can see, for a soft electron spectrum, $p_e\gtrsim 3$, given the physical constraints of 
Eqs. \ref{constraint1} and \ref{constraint2}, the above equation generally holds for all proton 
energies that significantly contribute to the radiative output. For hard electron spectrum, 
$p_e\lesssim 3$, the low-energy protons may be subject to dominant pion-production losses. However, the total radiative output of these low-energy protons will
be negligible compared to the output by ultrarelativistic protons ($\gamma_p \gtrsim 10^7$) and
can therefore be safely neglected.

Assumption 2 requires that the $\gamma\gamma$ optical depth satisfies $\tau_{\gamma \gamma} 
(\epsilon_{1}) < 1$. This implies a minimum Doppler factor of 
\begin{equation}
    \delta_{D} > \sqrt[6]{\frac{\sigma_{T} d_{L}^{2} f_{\epsilon^{obs}}^{pk} (1 + z)^{2} 
    \epsilon_{1}^{obs}}{4 m_{e} c^{4} t_{v}}}
    \label{deltaconstraint1}
\end{equation}
\citep{Dondi95}, where $d_{L}$ represents the luminosity distance of the source, $\epsilon_{1}^{obs}$ 
represents the highest observed energy of $\gamma$-ray photons, $f_{\epsilon^{obs}}^{pk}$ represents the 
observed flux of target photons and $t_{v}$ represents the variability time scale. The energy of the 
observed target photons in terms of the observed $\gamma$-ray photon energy is given by 
$\epsilon^{obs} = 2 \delta_{D}^{2}/[(1+z)^{2} \epsilon_{1}^{obs}]$. The observed flux at energy 
$\epsilon$ can be written in terms of the synchrotron photon field in the comoving frame of the jet,
\begin{equation}
    f_{\epsilon} = \epsilon F_{\epsilon} = \frac{\delta_{D}^{4} m_{e} c^{2} V_{b} \epsilon^{2} \ 
    n_{ph}(\epsilon)}{4 \pi d_{L}^{2} t_{lc}}
\end{equation}
where $t_{lc}$ is the light crossing time scale and $V_{b}$ represents the comoving volume of the emission 
region. High energy $\gamma$-rays of blazars typically peak around $\epsilon_{1}^{obs} \sim 1000$. The characteristic
energy of target photons for pair-production in an FSRQ, such as 3C 279, is then 
$\epsilon^{obs} = 2 \delta_{D}^{2}/[(1+z)^{2} \epsilon_{1}^{obs}] \sim 0.33$, which is in the hard X-ray band. 
This suggests that proton synchrotron emission represents the primary target photon field for pair-production. 
Assuming the target photon field is in the form of a power law, the constraint of Eq. (\ref{deltaconstraint1})
can be rewritten as
\begin{equation}
    \delta_{D} > \frac{1}{6}\sigma_{T} n_{ph}^{0}R \epsilon^{1 - \alpha} (1 + z)
\end{equation}
Again we apply the delta approximation for the synchrotron power of protons \citep{Boettcher12} and assuming 
a steady state proton distribution in the form of a power law, the constraint can then be rewritten in terms 
of the hadronic model parameters,
\begin{equation}
    \frac{\delta_{D}}{1+z} \gtrsim 10^{-51}(10^{-17})^{p_{p} - 2} R^2 B^{p_p} n_{0,p}\gamma_{p, max}^{p_{p}-1}
\end{equation}
where $R$ is in units of cm, $B$ in units of G, and $n_{0, p}$ in units of cm$^{-3}$. With the physical 
constraint in Eq. \ref{constraint2} and a typical Doppler factor of $\sim 20$, the above constraint is 
satisfied for all parameter combinations employed in this study.

Additionally, as it has been shown in many hadronic fittings \citep[e.g.,][]{Boettcher13,Cerruti15}, 
Compton scattering generally does not make a substantial contribution due to the large magnetic field, 
hence we will also neglect this effect here.

\subsection{Comparison with One-zone Code \label{compare}}

In order to verify the validity of our multi-zone hadronic radiation transfer approach, we compare the 
results of the one-zone code of \cite{Diltz15} to the results obtained with 3DHad+3DPol. We consider 
two sets of parameters for the quiescent state, Set 1 and 2, as listed in Table \ref{quiescent}. These parameters refer to the pre-flare equilibrium state, where all cells are characterized by the same set of parameters.
The difference between these two parameter sets is that Set 1 has particle evolution time scales 
generally larger than the light crossing time scale, while in Set 2 the light crossing time scale
is generally the longest relevant time scale. Both parameter sets obey the constraints derived above. 
In order to examine the flare features, we change the proton injection density rate for the 
entire emission region after equilibrium has been achieved: for Set 1, we choose
$\dot{u}_{p,inj}= 1.2\times 10^{-2} erg\,s^{-1}cm^{-3}$; for Set 2, 
$\dot{u}_{p,inj}= 1.5\times 10^{-5} erg\,s^{-1}cm^{-3}$.

\begin{figure}[ht]
\centering
\includegraphics[width=\linewidth]{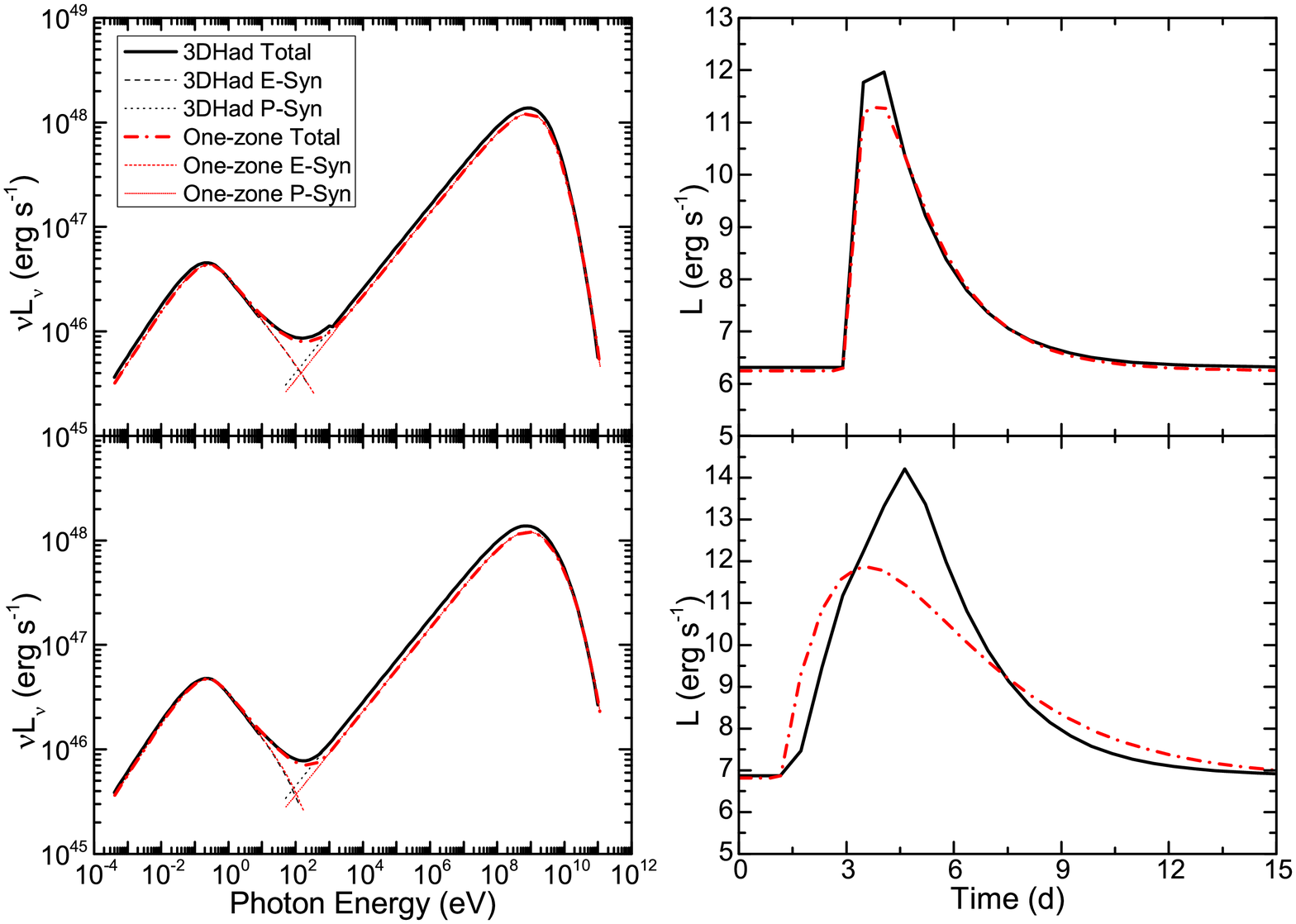}
\caption{Comparison of 3DHad and the one-zone hadronic code of \cite{Diltz15}. Left: Quiescent SEDs. 
Black curves show the 3DHad total SED (thick solid), as well as the electron synchrotron (dashed) and 
the proton synchrotron (dotted) individual contributions. Red lines show the one-zone code total (thick 
dashed-dotted), electron (short dashed), proton (short dotted). Right: Light curves. Black solid lines 
show the 3DHad output, while red dashed-dotted lines show the output of the one-zone code. Upper panel: 
parameter Set 1 with additional proton injection as described in Section \ref{compare}. Lower panel: 
parameter Set 2 with additional proton injection.
\label{comparison}}
\end{figure}

Since the particle evolution time scales are longer than the light crossing time scale in Set 1, we 
expect that the light curves from the one-zone code and 3DHad+3DPol should appear similar. For Set 2, 
however, since 3DHad+3DPol explicitly includes the LTTEs, we expect that the light curve will appear 
more symmetric in time than calculated with the one-zone code, which does not include LTTEs. 
Fig. \ref{comparison} presents the results. While minor differences probably due to the different 
geometry and the formulas are noticeable, the results generally meet our expectation. As a result, 
we conclude that 3DHad+3DPol is in agreement with the corresponding one-zone model for an appropriate
choice of geometry.

\begin{table}[ht]
\scriptsize
\centering
\begin{tabular}{|l|c|c|}\hline
Parameters                                          & Set 1  & Set 2        \\ \hline
Bulk Lorentz factor $\Gamma$                        & $20.0$ & $20.0$       \\ \hline
Orientation of LOS $\theta_{obs}$ $(^{\circ})$      & $90$   & $90$         \\ \hline
Radius of the emission region $R$ $(10^{16}cm)$     & $1.0$  & $9.0$        \\ \hline
Height of the emission region $Z$ $(10^{16}cm)$     & $1.33$ & $12.0$       \\ \hline
Acceleration time scale $t_{acc}$ $(10^6 s)$    & $8.2$  & $2.9$        \\ \hline
Escaping time scale $t_{esc}$ $(10^6 s)$        & $2.0$  & $1.5$        \\ \hline
Background injection electron density rate $\dot{u}_{e,inj}$ $(erg\,s^{-1}cm^{-3})$  & $2.8\times 10^{-7}$ &  $3.9\times 10^{-10}$      \\ \hline
Background injection electron minimum energy $\gamma_{e,min}$    & $100$    & $80$     \\ \hline
Background injection electron maximum energy $\gamma_{e,max}$    & $10000$  & $3500$   \\ \hline
Background injection electron spectral index $p_e$               & $2.8$    & $2.8$    \\ \hline
Background injection proton density rate $\dot{u}_{p,inj}$ $(erg\,s^{-1}cm^{-3})$    & $3\times 10^{-3}$ &  $1.9\times 10^{-6}$      \\ \hline
Background injection proton minimum energy $\gamma_{p,min}$  & $1$   & $1$  \\ \hline
Background injection proton maximum energy $\gamma_{p,max}$  & $5\times 10^8$   & $3\times 10^8$  \\ \hline
Background injection proton spectral index $p_p$               & $2.2$    & $2.2$    \\ \hline
Helical magnetic field $B$ $(G)$                    & $50.0$ & $80.0$       \\ \hline
Magnetic pitch angle $\theta_B$ $(^{\circ})$        & $45$   & $45$         \\ \hline
\end{tabular}
\caption{Summary of model parameters in the quiescent state. Except for the Lorentz factor, which is in 
the observer's frame, all parameters are in the comoving frame of the emission region. The time resolution 
is always identical to the typical light crossing time of a zone. Due to the small size of the emission
region in parameter Set 1, it is computationally expensive to obtain high time resolution. The resulting 
light curves and polarization signatures, however, due to the implicit Euler method employed
to solve the Fokker-Planck equations, we still obtain stable solutions for relatively large time steps
(Figs. \ref{case1a} to \ref{case1d}).
\label{quiescent}}
\end{table}

\section{Model Setup \label{model}}

The purpose of this paper is to study the general radiation and polarization signatures of hadronic 
blazar models. To show the most generic features of such models, we employ a simple model setup with 
the least physical assumptions. We assume that a cylindrical emission region travels relativistically 
in a straight trajectory along the jet, when it encounters a flat stationary disturbance, resulting in 
a flare. While we are observing blazars at a small observing angle $\theta^{\ast}_{obs}$ along the jet 
in the observer's frame, due to relativistic aberration, the observing angle $\theta_{obs}$ is much larger 
in the comoving frame of the emission region. Specifically, if $\theta^{\ast}_{obs} = 1/\Gamma$, where 
$\Gamma$ is the Lorentz factor of the emission region in the observer's frame, then $\theta_{obs} = 90^{\circ}$
for $\Gamma \gg 1$. As observations frequently suggest $\theta^{\ast}_{obs} \sim 1/\Gamma$, we will 
choose $\theta_{obs} = 90^{\circ}$. In this case, the Doppler factor is 
$\delta \equiv ( \Gamma [1 - \beta_{\Gamma} \cos\theta_{obs}^{\ast}] )^{-1} = \Gamma$.

In the comoving frame, the emission region is pervaded by a helical magnetic field. For this preliminary 
study, we will not add any turbulent field component. While a turbulent magnetic field is likely to dominate
the polarization fluctuations occurring mostly in the quiescent state \citep{Marscher14}, during major flares 
the polarization signatures appear more systematic, indicating a deterministic process 
\citep[e.g.,][]{Abdo10,Blinov15,Kiehlmann16}. \cite{ZHC15} have explicitly demonstrated that in such cases, 
the addition of a turbulent field component indeed yields better fit to the observational data, but the 
general trends of radiation and polarization signatures are similar to a purely helical field.

The disturbance will propagate through the emission region in the comoving frame. The zones affected by the 
disturbance will have different physical conditions from the initial state. After the disturbance moves out
of a given zone, the zone will revert to its initial conditions. We point out that while the physical 
conditions such as the stochastic acceleration and the nonthermal particle injection can reasonably 
return to the quiescent state after the passage of the disturbance, this is not necessarily the case
for the magnetic field strength and topology. However, the polarization signatures are frequently 
observed to quickly return to the initial values even after major variations such as PA swings 
\citep[e.g.,][]{Abdo10,Morozova14,Blinov15,Blinov16}, indicating the restoration of the magnetic 
field. \cite{ZHC16} have shown that this restoration is only possible when there is substantial 
magnetic energy compared to the plasma kinetic energy in the emission region. In most hadronic 
models, the magnetic energy is comparable to or stronger than the kinetic energy \citep[e.g.,][]{Boettcher13}. 
In particular, the parameter sets we will use in the following satisfy this condition. 

\begin{figure}[ht]
\centering
\includegraphics[width=\linewidth]{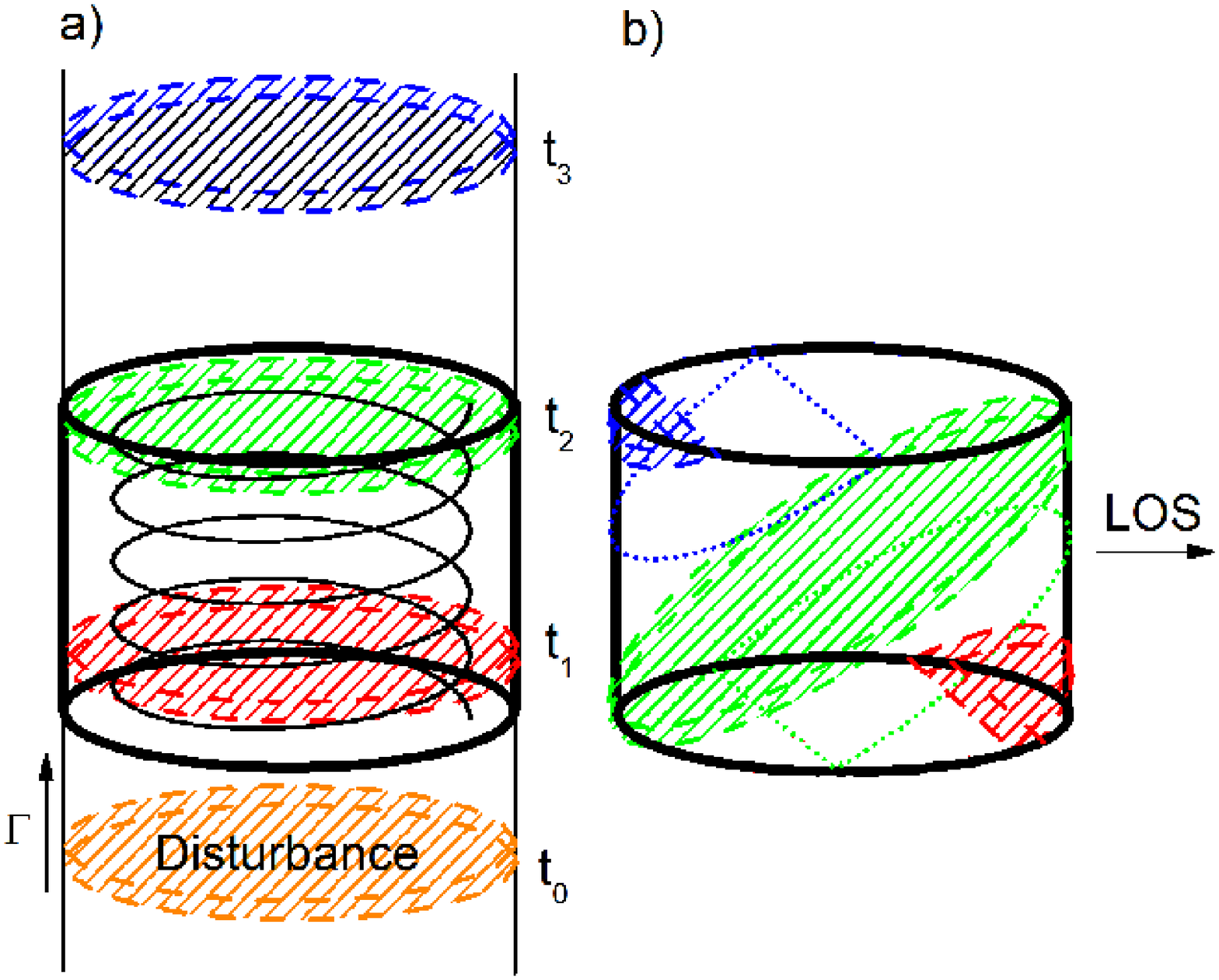}
\caption{Sketch of the model and LTTEs. Left: a disturbance propagates through the emission region pervaded 
by a helical magnetic field in its comoving frame. Red, green, and blue colors denote the location of the 
disturbance at approximately entering ($t_1$), leaving the emission region ($t_2$), and some time after 
leaving the emission region ($t_3$). Right: the corresponding flaring region at the $t_1$ to $t_3$ at equal
photon-arrival times at the observer. Dashed shaded regions are the active region; the region between the 
dashed shaded region and the dotted shape is the evolving region.
\label{LTTE}}
\end{figure}

Due to the LTTEs and the chosen $\theta_{obs}$, although the disturbance is flat, the observed 
``flaring region'' will appear different. Fig. \ref{LTTE} shows a sketch of our model, especially, 
the shapes of the flaring regions when the disturbance propagates through various locations in the 
emission region. The flaring region is composed of an ``active region'' with a slanted, ellipsoidal 
shape due to LTTEs, and an ``evolving region'' which is due to the slow evolution of protons. The 
zones outside the flaring region are termed as the ``quiescent region''. The impact of the LTTEs
on the polarization signatures has been discussed in detail in \cite{ZHC14,ZHC15}.

The parameter Sets 1 and 2 described in the previous section characterize the quiescent states for the 
following studies. We choose the same size of the disturbance in all the following case studies, which 
is $0.25$ times the length of the cylindrical emission region, so that the disturbance propagation time 
scale in a specific zone is $t_{\rm dp}=0.25 t_{\rm lc}$. We consider four scenarios in the active 
region that give rise to flares due to the disturbance:
\begin{enumerate}[a.]
\item Magnetic energy dissipation, where the magnetic field strength will decrease and its topology 
      will change, along with additional particle injection;
\item Magnetic compression, where the magnetic field strength will increase and change its topology;
\item Enhanced particle injection for both electrons and protons;
\item Enhanced stochastic acceleration, where the stochastic acceleration time scale becomes shorter.
\end{enumerate}
The flaring parameters for these scenarios are listed in Table \ref{flare}.
Since the flaring mechanisms are so different, in order to facilitate direct comparison, we choose 
the flaring parameters so that they result in approximately equal amplitudes of $\gamma$-ray flares. 
Also, we define similar epochs in the $\gamma$-ray light curves, approximately at the quiescent state, 
before the flare peak, at the peak of the flare, and after the peak. We point out that the low-energy 
(electron-synchrotron) light curves can look very different from the $\gamma$-ray light curves due to 
the drastically different radiative cooling time scales of protons and electrons.

\begin{table}
\scriptsize
\centering
\begin{tabular}{|l|c|c|c|c|c|c|}\hline
Parameters                            & Case 1a             & Case 2a              & Case 1b             & Case 1c             & Case 2c           & Case 1d     \\ \hline
$t_{acc,d}$ $(10^6 s)$                & --                  & --                   & --                 & --                  & --                  & $0.51$      \\ \hline
$\dot{u}_{e,inj,d}$ $(erg\,s^{-1}cm^{-3})$  & $3.0\times 10^{-6}$ & $4.1\times 10^{-9}$  & --                 & $1.6\times 10^{-6}$ & $2.2\times 10^{-9}$ & --          \\ \hline
$\dot{u}_{p,inj,d}$ $(erg\,s^{-1}cm^{-3})$  & $5.5\times 10^{-2}$ & $9.8\times 10^{-6}$  & --                 & $5.5\times 10^{-2}$ & $9.6\times 10^{-6}$ & --          \\ \hline
$B_d$ $(G)$                           & $36.6$              & $58.6$               & $136.6$           & --                  & --                  & --          \\ \hline
$\theta_{B_d}$ $(^{\circ})$           & $75$                & $75$                 & $75$               & --                  & --                  & --          \\ \hline
\end{tabular}
\caption{Summary of the model parameters at the disturbance. Only the parameters that are varied in the 
case studies are listed here. All parameters have the same meaning as in Table \ref{quiescent}, except 
for the subscript $d$ which denotes the parameters at the disturbance. \label{flare}}
\end{table}

\section{Synchrotron Cooling and LTTEs \label{result1}}

In this section, we will study Scenario a, magnetic energy dissipation to illustrate the effect of the relation between synchrotron cooling time scales and LTTEs. \cite{Guo16} have performed 
comprehensive PIC simulations to show that both electrons and protons can be effectively accelerated 
during magnetic reconnection events. As the detailed simulation of a reconnection event is beyond the 
scope of this paper, we simply assume that the particle injection is enhanced and the magnetic field 
is weakened in the disturbance, and keep the particle injection index the same as in the quiescent state. 
We have not included any thermal radiation contributions to the multiwavelength emission, such as the 
big blue bump typically seen in the flat spectrum radio quasars, or the host galaxy. In this way, the 
difference in the time-dependent radiation and polarization signatures mostly originates from the intrinsic 
hadronic physics. Proton synchrotron cooling is much slower than that for electrons. Due to the strong 
magentic field, the electron cooling time scale ($t_{\rm ec}$) is generally shorter than the light crossing 
time scale ($t_{\rm lc}$). Moreover, as is shown in Eq. \ref{constraint1}, the size of the emission region 
in the hadronic model has an upper limit due to the magnetic-flux constraint. Thus in most applicable 
situations, the proton cooling time scale ($t_{\rm pc}$) is comparable to or longer than $t_{\rm lc}$. In 
the following, we will demonstrate that this time scale relation, $t_{\rm ec} < t_{\rm lc} \lesssim t_{\rm pc}$, 
which is largely intrinsic to the hadronic model without any strong parameter dependence, is governing 
the general shape of the light curves and polarization signatures.

\subsection{Case 1a}

\begin{figure}[ht]
\centering
\vspace{-2cm}
\includegraphics[width=\linewidth]{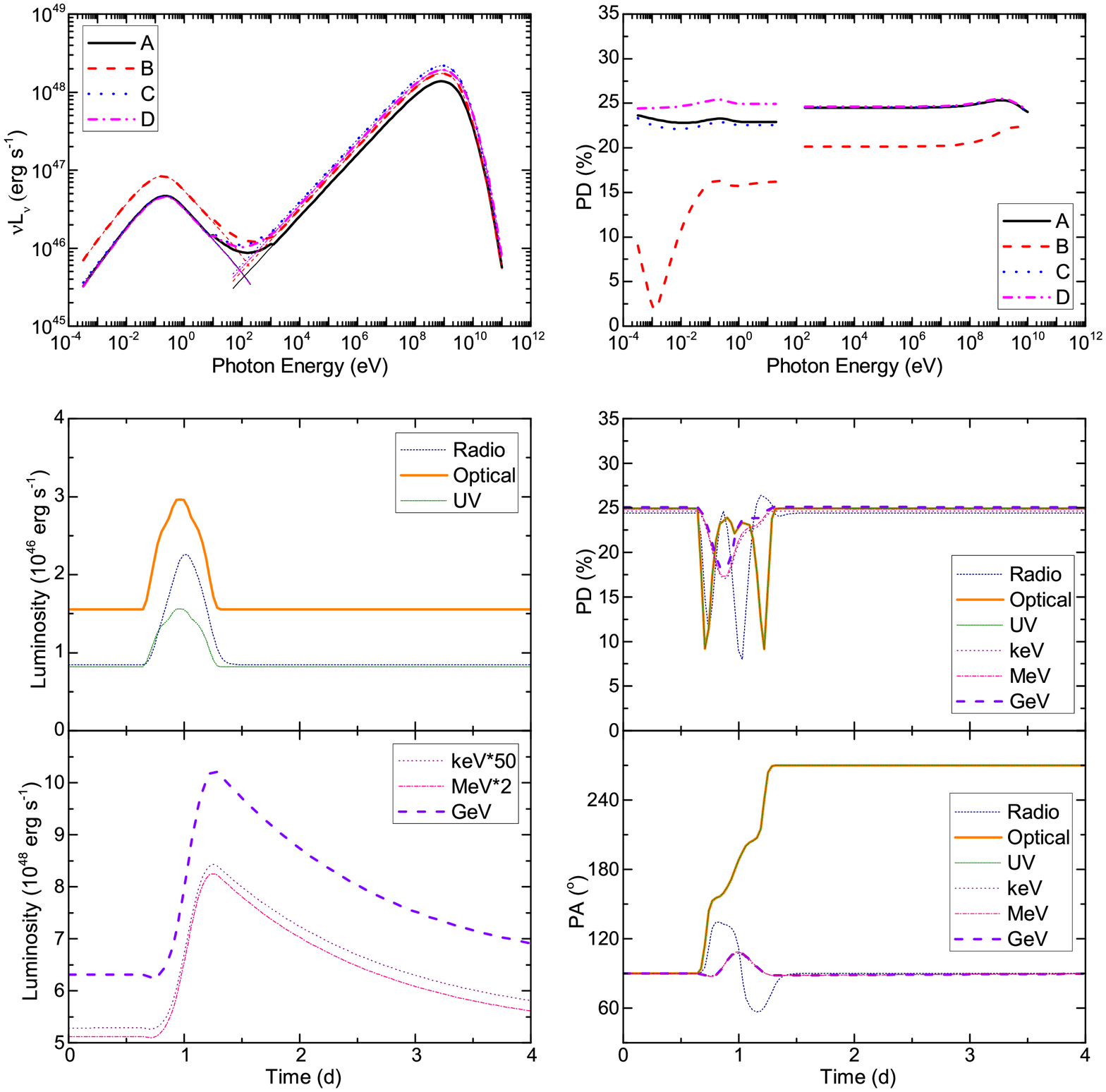}
\caption{Case 1a. Upper left: snap-shot SEDs approximately in the quiescent state (black solid, A), 
shortly before the flare peak (red dashed, B), at the flare peak (blue dotted, C), and after the flare 
peak (magenta dashed-dotted, D). Thin curves show the individual contributions from electron synchrotron 
and proton synchrotron. Upper right: snap-shot polarization degree vs photon energy. Curves are chosen 
at the same epochs as the SEDs. Middle and lower left: multiwavelength light curves chosen at radio 
(30 to 300 GHz, navy short-dashed), optical (1.8 to 3.2 eV, thick orange solid), UV (3.3 to 6.2 eV, 
olive short-dotted), X-ray (60 to 200 keV, purple dotted), MeV $\gamma$-ray (5 to 200 MeV, AdEPT, 
pink dashed-dotted), and GeV $\gamma$-ray (20 MeV to 300 GeV, {\it Fermi}-LAT, thick violet dashed) 
bands. Due to the large bandwidth of the GeV light curve, it collects a much higher total luminosity 
than the keV and MeV bands. Hence we manually boost those two bands by a fixed number to allow us to
show them in the same figure. Middle and lower right: multiwavelength PD and PA vs time. Bands are 
chosen the same as light curves.
\label{case1a}}
\end{figure}

This case refers to a small emission region (baseline parameter set 1) with magnetic energy dissipation 
(flaring scenario a) and is illustrated in Figure \ref{case1a}. Since we do not vary the power-law index 
for the enhanced particle injection, the SEDs (Fig. \ref{case1a} upper left) generally keep the same 
spectral indices during flares. In the quiescent state, the PD vs photon energy (Fig. \ref{case1a} 
upper right) displays minor fluctuations across the entire spectrum. However, during the flare, we 
notice considerable spectral PD variations. These are shown in more detail in the time evolution of 
radiation and polarization signatures (middle and bottom panels of Fig. \ref{case1a}).

Before we move to the emission evolution, we first take a look at the particle evolution. 
Fig. \ref{particle} presents the electron (upper left) and proton (upper right) spectral evolution 
for Case 1a. In the electron evolution, owing to the short $t_{\rm ec}$, after the disturbance 
leaves a certain zone, most electrons only take about a disturbance propagation time scale $t_{\rm dp}$ 
to revert to the pre-flare equilibrium. However, the lowest-energy electrons have longer synchrotron 
cooling timescales, thus they take longer (up to $\sim 4 t_{\rm dp}=1 t_{\rm lc}$) to revert to the quiescent state. 
Therefore, the evolving region for the electrons is comparable to the active region, except for the 
lowest-energy electrons where it is moderately larger. The proton evolution appears very different. 
In view of the much longer cooling time scale $t_{\rm pc}$, after about $20 t_{\rm dp}$, the proton 
synchrotron contribution from the evolving region no longer dominates the active region, and it takes 
about $60 t_{\rm dp}$ (i.e., about $15 t_{\rm lc}$) to evolve back to a state close to the quiescent 
equilibrium. Hence for protons the evolving region overwhelms the active region.

These features are clearly reflected in light curves and polarization variations. For the low-energy 
component, since the evolving region for electrons is relatively small, the LTTEs will be dominating. 
Thus the flaring region on the right side of the sketch in Fig. \ref{LTTE} is generally similar to 
that on the left side. Therefore, the light curves from radio to UV appear generally symmetric in 
time (Fig. \ref{case1a} middle left). However, due to the slightly longer $t_{\rm ec}$ for the 
lowest-energy electrons, which are responsible for the radio emission, the radio peaks a little 
bit later than the optical and UV. Unlike the light curves, which only depend on the luminosity, 
the polarization signatures are also affected by the PD and PA in each zone. As the magnetic field 
structure varies from the active region to the evolving and the quiescent region, even in the PD 
and PA of the optical and UV bands we find a small degree of asymmetry in time. We notice that 
both the optical and UV bands exhibit a PA swing and significant PD variations (Fig. \ref{case1a} 
middle and lower right). This is consistent with the results of \cite{ZHC14,ZHC15}, where these
effects are discussed in detail. 

The proton evolving region dominates the high-energy emission. All light curves peak 
considerably later than the low-energy light curves; in particular, we find long cooling tails 
in the high-energy light curves, which strongly extend the flare duration (Fig. \ref{case1a} lower 
left). In the polarization signatures, we only find small changes in both PD and PA (Fig. \ref{case1a} 
middle and lower right), due to the strong contamination from the evolving region, which has the same magnetic topology as the quiescent region. Specifically, at 
the beginning of the flare, as the evolving region is very small, the PD shows a relatively large 
drop. After that the evolving region becomes dominant, hence the polarization gradually recovers 
its initial state. When the active region has completely moved out, approximately at the same time 
when the low-energy flare stops, the high-energy polarization signatures appear largely identical 
to the quiescent state.

\subsection{Case 2a}

\begin{figure}[ht]
\centering
\includegraphics[width=\linewidth]{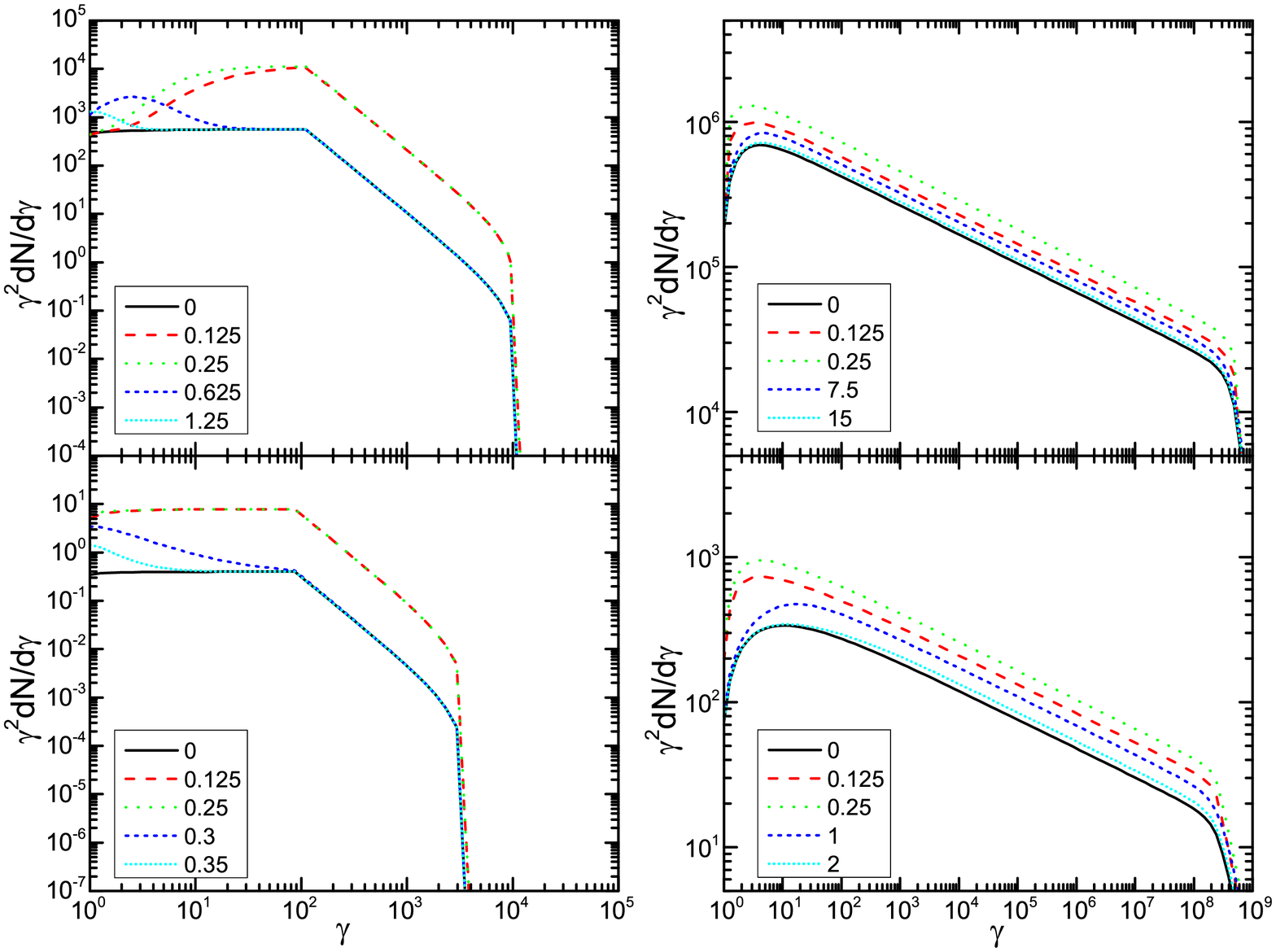}
\caption{Particle spectra for Case 1a and 2a. Left: electron spectra at various epochs. 
Right: proton spectra at various epochs. Upper: particle spectra for Case 1a. Lower: particle 
spectra for Case 2a. The particle spectra are chosen at various epochs in units of light 
crossing time scales ($t_{\rm lc}$) in both cases, approximately at the quiescent state (0), 
in the middle of the disturbance (0.125), as the disturbance leaves the zone (0.25), and at 
later times.
\label{particle}}
\end{figure}

The major difference of Case 2a (baseline parameter set 2) compared to Case 1a lies in the longer light crossing time, $t_{\rm lc}$. 
Again we examine the particle evolution (Fig. \ref{particle} lower panel). When the disturbance moves 
out of a certain zone, most electrons revert to the quiescent equilibrium immediately, except for the 
electrons responsible for the radio emission, which take up to about $0.5 t_{\rm dp}$ to recover. 
Thus, the evolving region is generally smaller than the active region. For protons, after the 
disturbance leaves, they continue to make a substantial contribution to the high-energy emission 
for $\sim 3 t_{\rm dp}=0.75 t_{\rm dp}$, although it takes $\sim 7 t_{\rm dp}$ to fully recover to equilibrium. 
Hence, the evolving region is moderately larger than the active region.

\begin{figure}[ht]
\centering
\includegraphics[width=\linewidth]{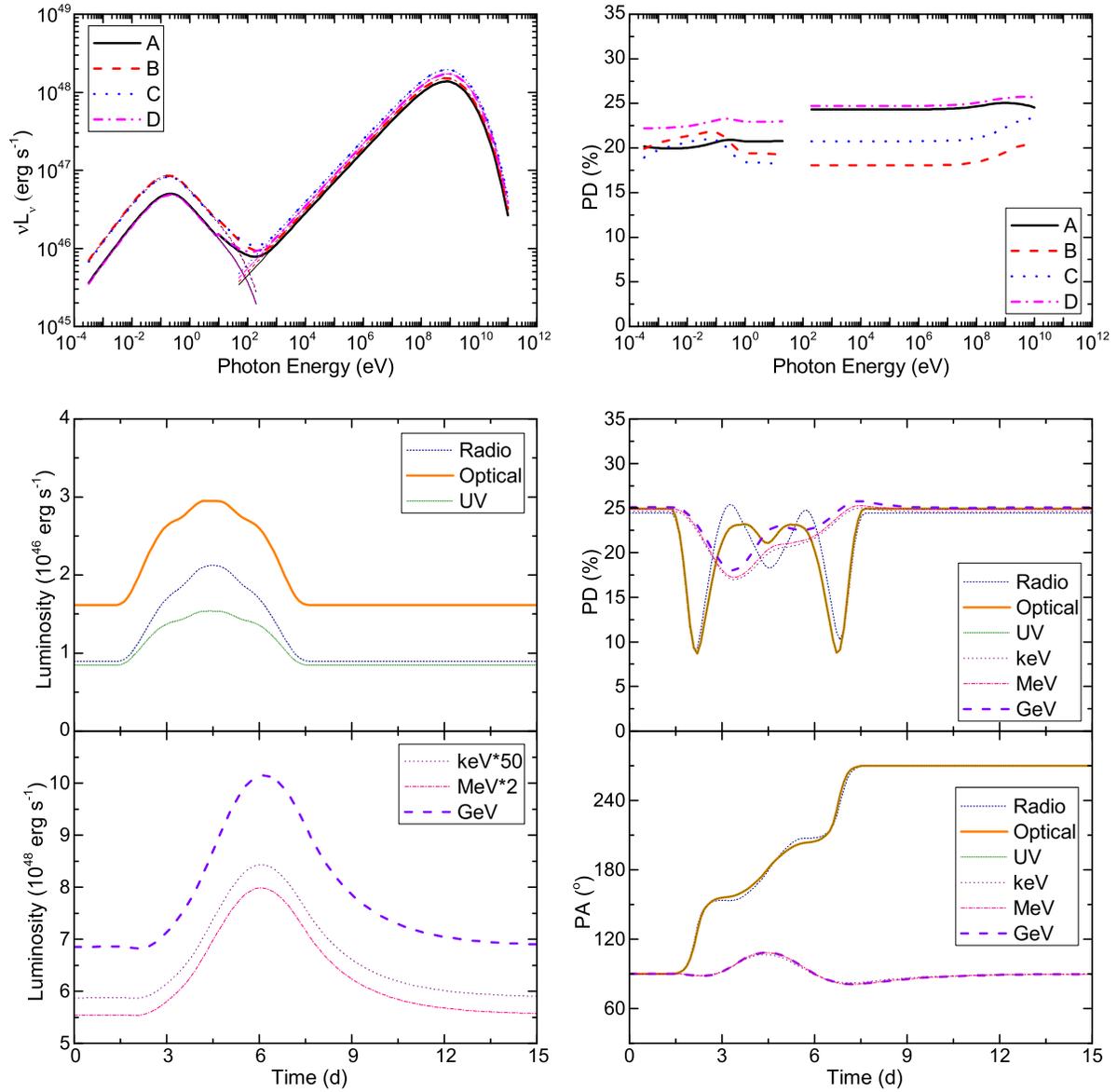}
\caption{Case 2a. Due to the larger emission region, the flare duration is longer. Otherwise, panels 
and line styles are the same as in Fig. \ref{case1a}.
\label{case2a}}
\end{figure}

Consequently, in the low-energy component, all light curves and polarization signatures appear 
symmetric in time without any noticeable delay (Fig. \ref{case2a} middle left); additionally, 
all bands display PA swings, although the radio polarization slightly diverges from the others 
(Fig. \ref{case2a} middle and lower right). In the high-energy component, the light curves appear 
generally symmetric in time, though they still peak later and the flares last longer than in the 
low-energy light curves (Fig. \ref{case2a} lower left). Nevertheless, the polarization contamination 
from the evolving region is still significant, thus there is no PA swing in the high-energy bands 
(Fig. \ref{case2a} middle and lower right). We notice that unlike the light curves, the high-energy 
polarization signatures are still synchronized with the low-energy flares and polarization variations: 
they all end approximately when the active region moves out.

To summarize this section, we find that the combined effects of synchrotron cooling and LTTEs will 
result in some interesting features in the hadronic models. First, the polarization signatures in the 
high-energy component are nearly identical from X-rays to $\gamma$-rays. Therefore,  X-ray and $\gamma$-ray 
polarimeters may both be able to measure hadronic signatures in the high-energy polarization. Additionally, 
in the quiescent state, if the electrons and protons reside in the same emission region, the high-energy 
polarization signatures should be generally identical to the low-energy component. 
Moreover, the low-energy light curves and polarization variations are generally symmetric in time, but 
the high-energy signatures are generally asymmetric. Also the high-energy flares generally peak later 
and last longer than the low-energy flares. The low-energy flares and polarization variations, as well 
as the high-energy polarization variations, are generally synchronized. Finally, while the low-energy 
polarization signatures may vary rapidly during flares, high-energy polarization signatures appear 
generally stable.

\section{Case Study of Alternative Flaring Scenarios \label{result2}}

In this section, we will present more case studies to further test the hadronic features listed in 
the previous section, and examine how the different flaring mechanisms may affect the radiation and 
polarization signatures. We notice that the radio emission from blazars is generally dominated by 
the large-scale jets instead of the local emission region, therefore, we will not show the radio 
signatures in the following. Moreover, the optical and UV bands appear identical; the same applies 
to the keV, MeV, and GeV bands. Thus in the following we will only take the optical and {\it Fermi}-LAT 
GeV band as representative for the low- and high-energy components, respectively, and term the GeV 
band simply as ``$\gamma$-ray''.

\subsection{Scenario b}

\begin{figure}[ht]
\centering
\includegraphics[width=\linewidth]{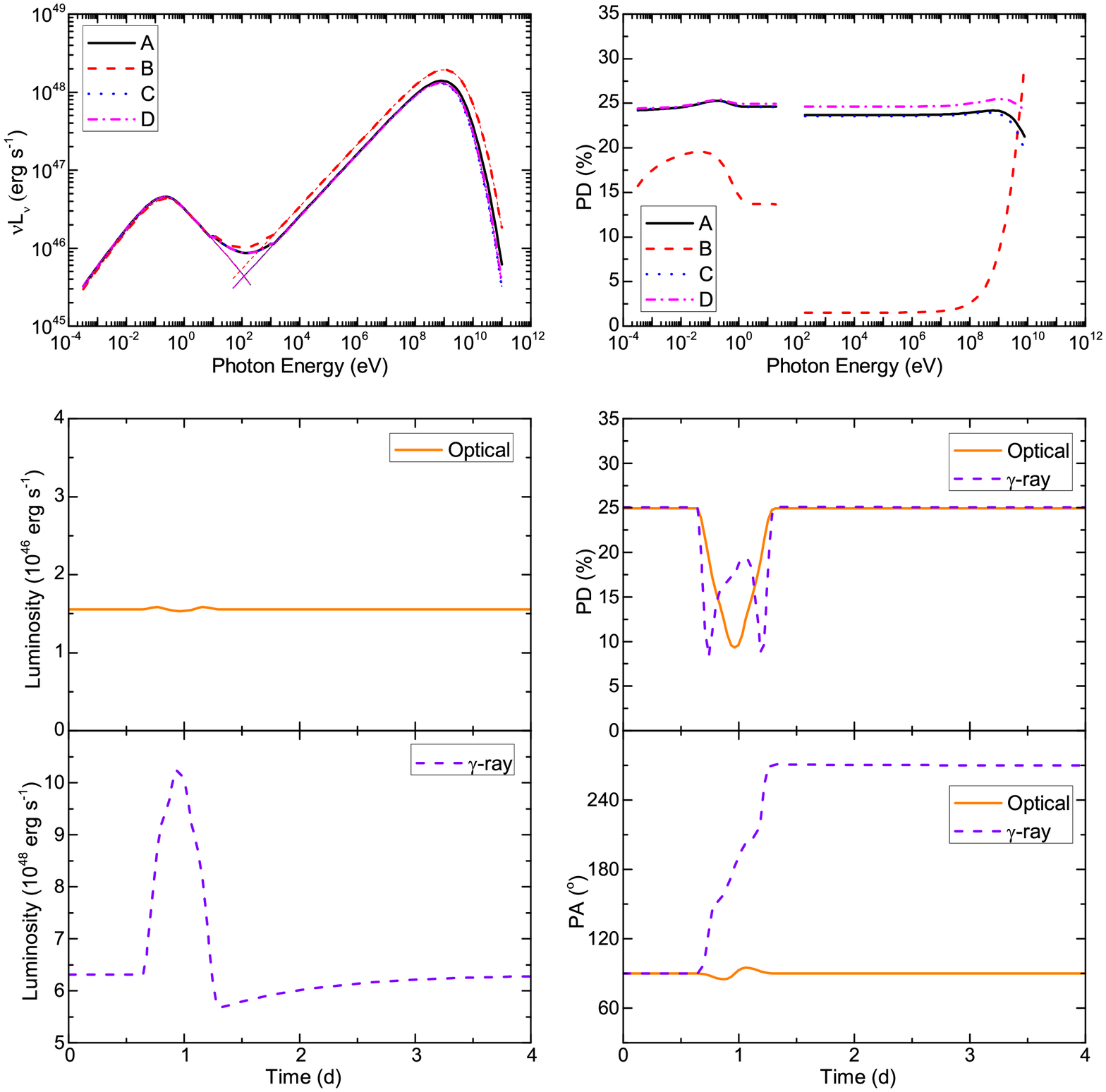}
\caption{Case 1b. Compared to Fig. \ref{case1a}, we removed the radio, UV, keV and MeV bands, and 
termed the GeV band as $\gamma$-ray. Otherwise, panels and line styles are the same as in Fig. \ref{case1a}.
\label{case1b}}
\end{figure}

We first look at Case 1b, compression of the magnetic field. In the presence of a strong magnetic 
field, most electrons are efficiently synchrotron cooled. This means that the electron synchrotron 
is already at maximal radiative efficiency. Consequently, the enhanced magnetic field will not boost 
the low-energy synchrotron flux, so that no flare is observed in the optical band (Fig. \ref{case1b} 
middle left). However, the active region possesses a dominant toroidal magnetic field topology, 
leading to a considerable drop in the PD (Fig. \ref{case1b} middle right). Nevertheless, the 
active region does not provide additional emission, thus the enhanced toroidal magnetic field 
alone is unable to trigger a PA swing (Fig. \ref{case1b} lower right).

On the other hand, the high-energy synchrotron component exhibits some interesting features. 
Due to the enhanced magnetic field in the active region, the proton cooling becomes faster, 
giving rise to higher flux. Additionally, after the disturbance moves out a certain zone, 
the proton spectrum has a lower normalization than in the quiescent state. Hence the evolving 
region actually provides less emission than the quiescent region, leading to a minor contamination 
in the emission signatures. This is clearly shown at the end of the flare (Fig. \ref{case1b} lower 
left), where the $\gamma$-ray light curve drops below the initial value. Also in the PD vs photon 
energy, we find a rapidly increasing PD tail at the highest energy, due to the exponential cooling 
cut-off in the proton spectrum (Fig. \ref{case1b} upper right). In this way, the radiation and 
polarization signatures during the flare are dominated by the active region. Therefore, the 
$\gamma$-ray light curve becomes generally symmetric in time, and there are significant and 
generally time-symmetric PD changes and a PA swing during the flare (Fig. \ref{case1b} middle, 
lower right).

In conclusion, the special properties in this scenario include an orphan $\gamma$-ray flare, and 
major polarization variations in both low- and high-energy components. In particular, there may 
be a $\gamma$-ray PA swing. Nonetheless, we want to emphasize that in the hadronic model, the 
magnetic field is very strong, and in most cases carries energy comparable to or even stronger 
than the plasma kinetic energy. In order to adequately compress the magnetic field, strong shocks 
are necessary, which are unlikely to happen in such a highly magnetized environment \citep{Komissarov11}. 
As a result, we suggest that this scenario is unlikely in practice. For the parameter Set 2, the 
magnetic energy is much stronger than the kinetic energy, further prohibiting this scenario. 
Thus we will not discuss Case 2b.

\subsection{Scenario c}

\begin{figure}[ht]
\centering
\includegraphics[width=\linewidth]{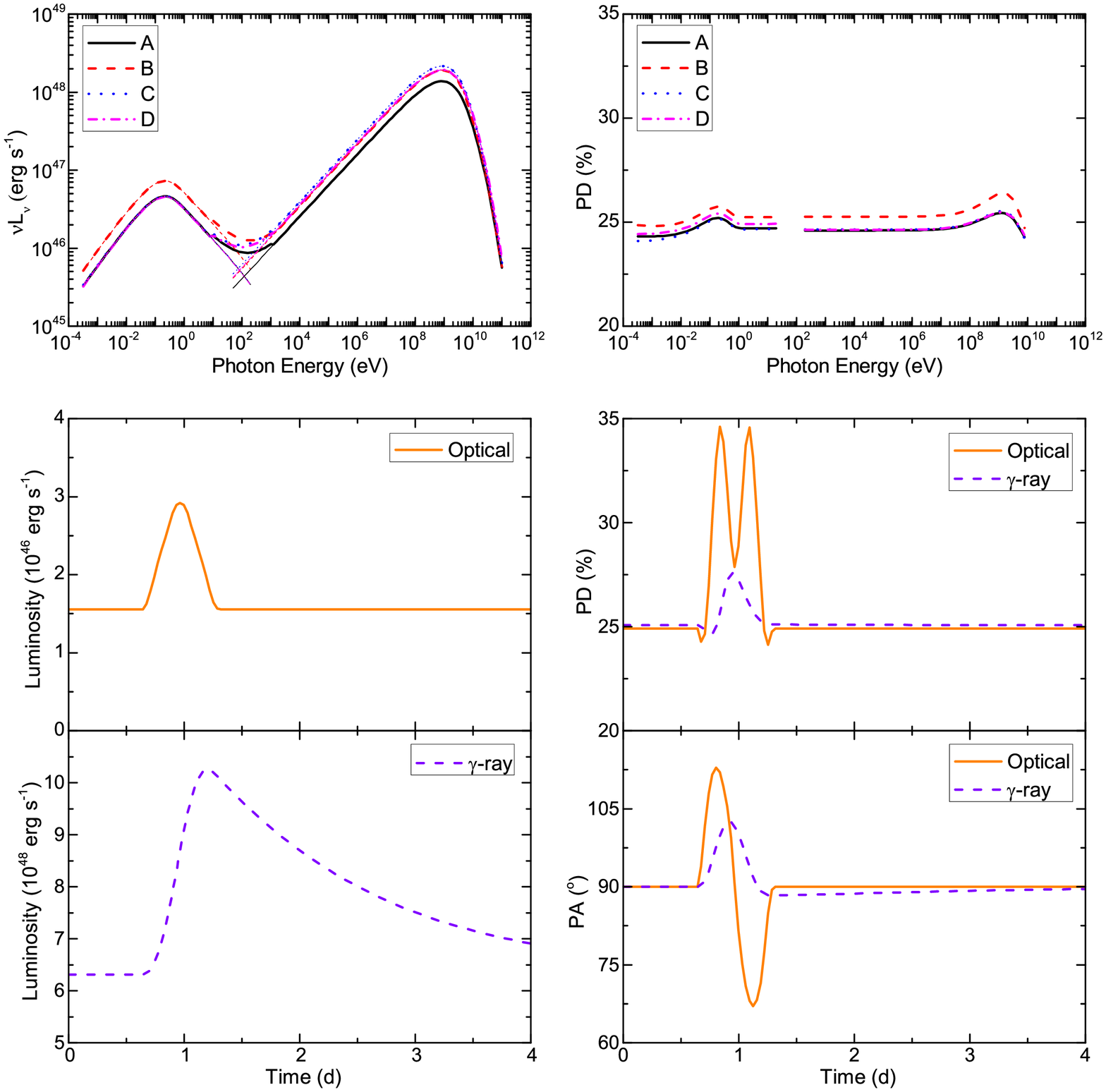}
\caption{Case 1c. Panels and line styles are the same as in Fig. \ref{case1a}.
\label{case1c}}
\end{figure}

\begin{figure}[ht]
\centering
\includegraphics[width=\linewidth]{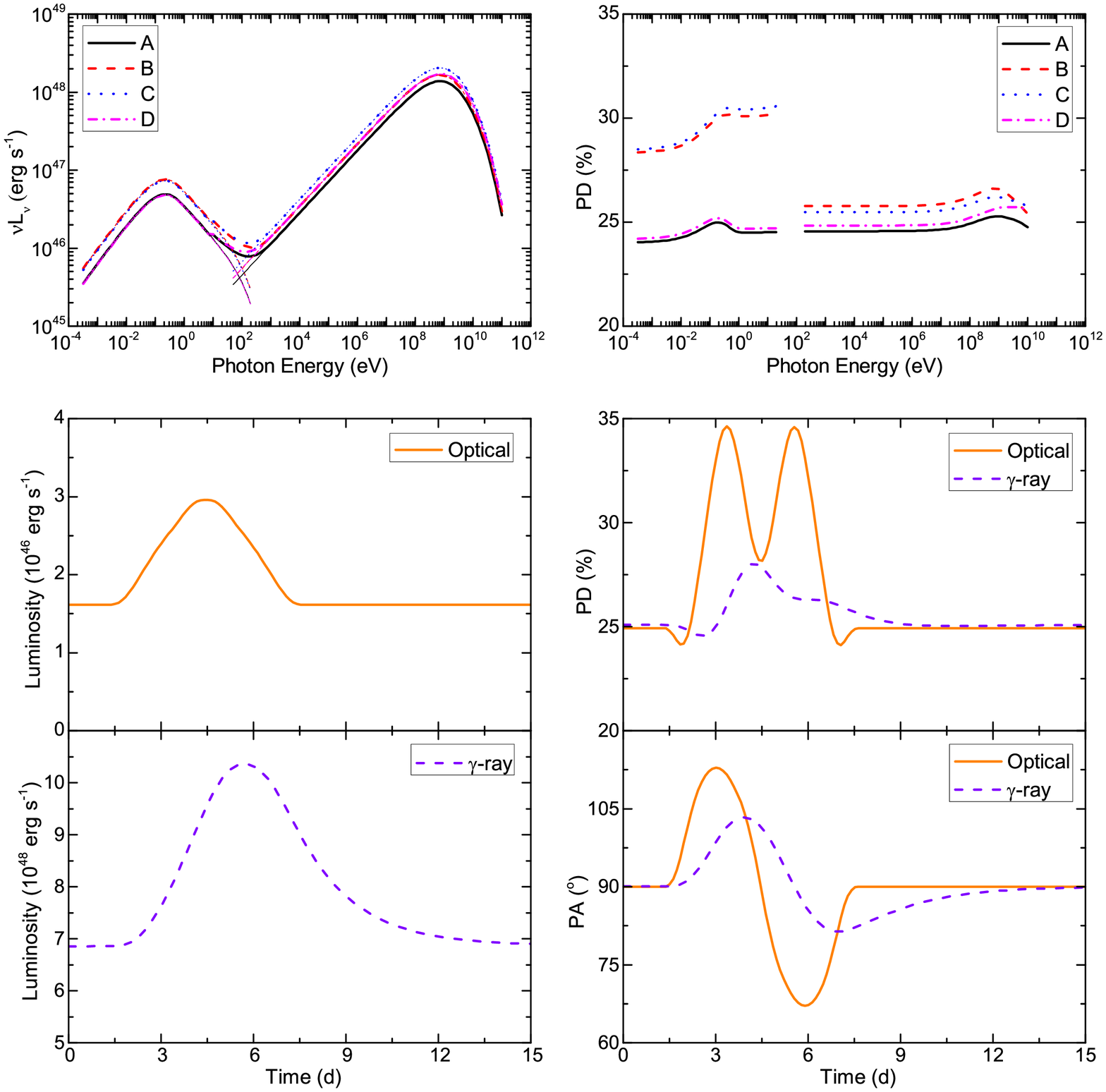}
\caption{Case 2c. Compared to Fig. \ref{case2a}, we removed the radio, UV, keV and MeV bands, and termed 
the GeV band as $\gamma$-ray. Otherwise, panels and line styles are the same as in Fig. \ref{case2a}.
\label{case2c}}
\end{figure}

This is the case of enhanced particle injection at the disturbance without changing the magnetic field.
Hence, the synchrotron cooling rates remain unchanged. The polarization variations in the optical band 
generally arise from the active region as it energizes up different parts of the emission region during 
its propagation (Fig. \ref{case1c} middle and lower right). This has been discussed in detail in \cite{ZHC14}.
On the other hand, owing to the large evolving region of the proton population, the high-energy 
polarization signatures again appear asymmetric in time and exhibit smaller variations than the 
low-energy ones. Compared to Case 1a, since the magnetic field topology is unchanged, we find
an increase in the PD instead (Fig. \ref{case1c} middle and lower right). For Case 2c, LTTEs 
dominate. As a result, both the $\gamma$-ray light curve and the PA variability appear generally 
symmetric in time (Fig. \ref{case2c} lower). However, we can still see in the PD that the large 
evolving region will substantially contaminate the PD from the active region, giving rise to an 
asymmetric time variation (Fig. \ref{case2c} middle right). In conclusion, Scenario c (enhanced
particle injection) results in similar features as Scenario a (particle energization by magnetic
energy dissipation).

\subsection{Scenario d}

\begin{figure}[ht]
\centering
\includegraphics[width=\linewidth]{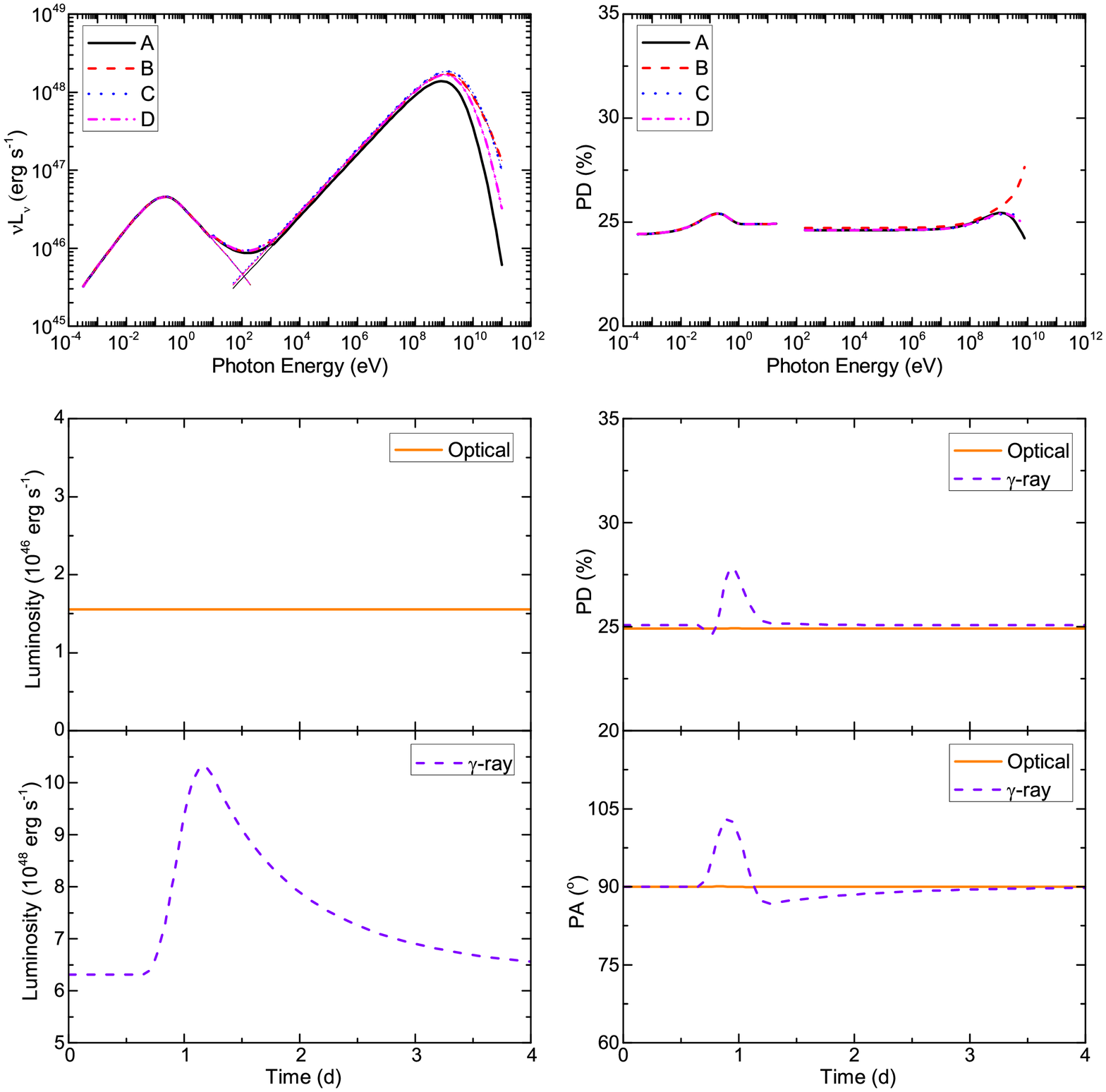}
\caption{Case 1d. Panels and line styles are the same as in Fig. \ref{case1a}.
\label{case1d}}
\end{figure}

We finally consider Scenario d, enhanced stochastic acceleration. We assume that the stochastic acceleration parameterized by $t_{acc}$, which represents the
wave-particle interaction with plasma waves in the turbulence, is enhanced due to the action of a
shock. In view of the very fast synchrotron 
cooling of electrons, even the shortened acceleration time scale is still too long to have a significant
impact on the electron distribution. Hence we observe featureless radiation and polarization signatures 
in the low-energy component (Fig. \ref{case1d} middle, lower right). In the high-energy component, the 
enhanced acceleration boost the protons to higher energy, so that the SED becomes harder above a few GeV 
(Fig. \ref{case1d} upper left). However, since the total particle injection rate is kept unchanged, no 
major change is detected at lower energies. Otherwise, the high-energy signatures are similar to Case 1c. 
The same applies to Case 2c, hence we will not discuss it in detail here.

We conclude that enhanced stochastic acceleration will result in a mild orphan $\gamma$-ray flare, 
similar to Case 1b. The differences are: phenomenologically, both low- and high-energy polarization 
signatures are stable in time, and there is no orphan flare in the X-ray band; physically, stochastic 
acceleration is due to magneto-hydrodynamic turbulence in the emission region, whose characteristics 
are likely to be altered by a passing shock. Therefore, we suggest that this case is more plausible 
than Scenario b.

\section{Discussions and Conclusion \label{discussion}}

In this paper, we have presented the first 3D multi-zone time-dependent lepto-hadronic blazar code, 
3DHad, describing a lepto-hadronic model in a parameter regime in which the high-energy
emission is dominated by proton synchrotron radiation. By coupling with the 3DPol code, we are able to 
derive the time-dependent flux and polarization signatures of this lepto-hadronic blazar emission model, 
including all LTTEs. Our work thus makes the first attempt to study the time-dependent lepto-hadronic 
multi-wavelength polarization signatures of blazar emission.

We have explicitly calculated the physical constraints for the hadronic model. Based on our estimates, 
if the Blandford-Znajek mechanism is responsible for powering the jet and providing the the magnetic field 
in the jet, the hadronic emission region cannot be very large due to the limited magnetic flux that the 
central black hole can provide. Therefore, the largest variability time scale in the observer's frame is 
unlikely to exceed a few days. Also, the high particle energy necessary for the lepto-hadronic scenario requires
extreme jet powers. These constraints would also suggest that 
UHE extragalactic neutrinos are unlikely to be attributed to blazars, as photo-pion production is negligible. If the lepto-hadronic polarization signatures derived here are indeed detected in future 
observations, the required extremely efficient particle acceleration, strong magnetic field, and high jet 
power will seriously challenge our current understanding of AGN jet formation.

We have demonstrated that the general time-dependent signatures of our proton-synchrotron dominated 
lepto-hadronic blazar model is dominated by the intrinsic time scale relations, namely, 
$t_{\rm ec} < t_{\rm lc} \lesssim t_{\rm pc}$. Through detailed parameter studies, we have identified 
the following time-dependent signatures of this model:
\begin{enumerate}
\item The time-dependent low-energy radiation signatures are generally symmetric in time, while the 
high-energy signatures are generally asymmetric;
\item The high-energy flares generally peak later and last longer than the low-energy flares;
\item An orphan flare in the high-energy component is possible;
\item The polarization signatures at various wavelengths within the high-energy component are generally similar;
\item In the quiescent state, if the low- and high-energy components are co-spatial, they share similar 
polarization degrees and angles.
\item While the low-energy polarization signatures may vary rapidly during flares, high-energy polarization 
signatures appear generally stable.
\item The time-dependent low-energy signatures and the high-energy polarization variations are generally 
synchronized with the disturbance propagation and the LTTEs. The high-energy flares, on the other hand, can 
last much longer due to the slow proton cooling.
\end{enumerate}
We suggest that these features can be tested with simultaneous multiwavelength observations, including 
future high-energy polarimetry.

We notice that the polarization signatures possess a strong dependence on the magnetic field evolution. 
Although we have demonstrated in Section \ref{model} that our assumptions on the magnetic field evolution 
are reasonable, our test cases are most likely an over-simplification of any actual physical scenario. 
However, our code can be easily coupled with first principle simulations, such as MHD, to constrain the 
magnetic field evolution, so that our polarization signatures in both low- and high-energy components are 
physically self-consistent.

\acknowledgments{HZ is supported by the LANL/LDRD program and by DoE/Office of Fusion Energy Science 
through CMSO. MB acknowledges support by the South African Research Chairs Initiative (SARChI) of the 
Department of Science and Technology and the National Research Foundation \footnote{Any opinion, finding 
and conclusion or recommendation expressed in this material is that of the authors and the NRF does not 
accept any liability in this regard.} of South Africa. Simulations were conducted on LANL's Institutional 
Computing machines.}

\clearpage

\end{document}